\documentclass{czjphys}         % for LaTeX 2e
\slacs{.4ex}
\begin{document}
\title{Anomalies in $\mathcal{PT}$-Symmetric Quantum Field Theory}

\authori{Kimball A. Milton}
\addressi{Department of Physics and Astronomy, University of Oklahoma,
Norman, OK 73019 USA}
\authorii{}     \addressii{}
\authoriii{}    \addressiii{}
\authoriv{}     \addressiv{}
\authorv{}      \addressv{}
\authorvi{}     \addressvi{}
\headauthor{K. A. Milton}
\headtitle{Anomalies in $\mathcal{PT}$-Symmetric Quantum Field Theory}
\lastevenhead{K. A. Milton:  
Anomalies in $\mathcal{PT}$-Symmetric Quantum Field Theory}
%%%%%%%%%%%%%% FOR EDITORIAL USE ONLY!!! %%%%%%%%%%%%%%%
\refnum{A}%\total{}\type{}
\daterec{XXX}    %;\\ final version }
\issuenumber{0}  \year{2003}
\setcounter{page}{1}
%\firstpage{1}
%\lastpage{000}
%\makefirsttitle
%%%%%%%%%%%%%%%%%%%%%%%%%%%%%%%%%%%%%%%%%%%%%%%%%%%%%%%%

\maketitle
\begin{abstract}
It is shown that a version of $\mathcal{PT}$-symmetric electrodynamics
based on an axial-vector current coupling massless fermions to the
photon possesses
anomalies and so is rendered nonrenormalizable.  An alternative theory
is proposed based on the conventional vector current constructed from massive
Dirac fields, but in which the $\mathcal{PT}$ transformation 
properties of electromagnetic fields are reversed.  Such a theory seems
to possess many attractive features.
\end{abstract}

\section{Introduction}     %\section*{Introduction}
In 1996 we proposed \cite{bm97} that a new class of quantum field
theories might exist, in which the Lagrangian need not be Hermitian,
yet the theory still might possess a positive spectrum.
 In these theories both
parity $\mathcal{P}$ and time-reversal $\mathcal{T}$ invariance were
violated, but the product $\mathcal{PT}$ symmetry was unbroken.
 Apparently, it is the presence of $\mathcal{PT}$ symmetry that
replaces Hermiticity in guaranteeing positivity of the spectrum.

In early papers we examined parity violation in scalar field theories with
interaction \cite{bm97}\begin{equation}
\mathcal{L}_{\rm int}=-g({\rm i}\phi)^N,\end{equation}
proved that the supersymmetry of theories possessing the superpotential
 \cite{bm98}
\begin{equation}
\mathcal{S}=-{\rm i}g({\rm i}\phi)^N
\end{equation}
was not broken by (nonperturbative) quantum corrections,  suggested
that a stable eigenvalue condition held in 
massless electrodynamics defined by an axial vector current \cite{bm99}
\begin{equation}
j_5^\mu=e\frac12\psi\gamma^0\gamma^\mu\gamma^5\psi,
\end{equation}
and showed that the Schwinger-Dyson equations for the theory \cite{bms00}
\begin{equation}
\mathcal{L}_{\rm int}=-g\phi^4
\end{equation}
possessed both perturbative and nonperturbative solutions.

\section{$\mathcal{PT}$-symmetric electrodynamics}

In this talk we will reconsider the $\mathcal{PT}$-symmetric version of
massless electrodynamics \cite{bm99}, 
which is described by the Lagrangian
\begin{equation}
\mathcal{L}=-\frac14F^{\mu\nu}F_{\mu\nu}-\frac12\psi\gamma^0\gamma^\mu
\frac1{\rm i}\partial_\mu\psi+e\frac12\psi\gamma^0\gamma^\mu\gamma^5 A_\mu\psi.
\end{equation}
Our conventions are the following:
$\gamma^0$ is antisymmetric and pure imaginary, $\gamma^0\gamma^\mu$ is 
symmetric and real, $\gamma^5=\gamma^0\gamma^1\gamma^2\gamma^3$ is 
antisymmetric and real, and $(\gamma^5)^2=-1$.  The fermion field $\psi$
is expected to be complex.

$\mathcal{L}$ possesses gauge invariance:
\begin{eqnarray}
A^\mu\to A^\mu+\partial^\mu\lambda,\quad
\psi\to{\rm e}^{{\rm i}e\gamma^5\lambda}\psi.
\end{eqnarray}
The gauge transformation on the fermion field is not a phase transformation,
but a scale transformation; nevertheless it leaves invariant the fermion
bilinears in the Lagrangian, and in the energy-momentum tensor.
Note that a mass term $\frac12m\psi\gamma^0\psi$
breaks this gauge symmetry.

We had expected (erroneously as we shall see)
the weak-coupling Feynman rules, for graphs with even numbers
of $\gamma_5$s to be the same as those in ordinary QED, except that 
$\alpha\to-\alpha$.  This is very intriguing, for it suggests that the
program of Johnson, Baker and Willey \cite{jbw} might succeed, for now
their eigenvalue condition for the fine structure constant reads,
in terms of the ``quenched'' beta function,
\begin{equation}
0=F_1(\alpha)=-\frac43\left(\frac\alpha{4\pi}\right)+4\left(\frac\alpha{4\pi}
\right)^2+2\left(\frac\alpha{4\pi}\right)^3-46\left(\frac\alpha{4\pi}
\right)^4+\dots,
\label{jbw}
\end{equation}
which displays all the coefficients calculated to date (remarkably integers).
Keeping two, three, and four terms in this series gives a sequence of
quite stable positive roots:
\begin{eqnarray}
\alpha_2&=&4.189,\\
\alpha_3&=&3.657,\quad (2,1) \mbox{ Pad\'e: } 3.590,\\
\alpha_4&=&4.110, (3,1) \mbox{ Pad\'e}.
\end{eqnarray}
\subsection{Erratum}

This is an appropriate point to acknowledge an error in Ref.~\cite{bm99}.
There it was stated that the eigenfunction condition in
conventional QED, obtained by replacing $\alpha\to-\alpha$ in Eq.~(\ref{jbw}),
possesses only the following successive positive roots of $F_1$:
\begin{eqnarray}
\alpha_3&=&13.872,\\
\alpha_4&=&3.969, \quad (1,2) \mbox{ Pad\'e: } 0.814, \quad(2,1) \mbox{ Pad\'e:
} 0.545.
\end{eqnarray}
These were nearly completely misstated.  The correct positive roots are
\begin{eqnarray}
\alpha_3&=&28.79,\\
\alpha_4&=&4.804, \quad  (2,1) \mbox{ Pad\'e of } F_1/\alpha: 0.545.
\end{eqnarray}  The conclusion, however, that the conventional QED
shows no sign of stability of the eigenvalue, is of course unaltered.

\subsection{Difficulties with this version of electrodynamics}
However, there is no Furry's theorem for this $\mathcal{PT}$-symmetric
 electrodynamics,
because the antisymmetrical charge matrix $q$ of ordinary QED
is replaced by the 
antisymmetrical $\gamma^5$ matrix: ($\psi$ is a Grassmann element)
\begin{equation}
j_\mu=\frac12\psi\gamma^0\gamma_\mu eq\psi\
\to\frac12\psi\gamma^0\gamma_\mu e\gamma^5\psi.
\end{equation}
Thus there is a three-photon triangle graph as shown in Fig.~\ref{fig1}.
\bfg[t]                     %instead of \begin{figure}[t]
\bc                         %instead of \begin{center}
\mbox{
\begin{picture}(200,100)
\multiput(100,0)(0,10){4}{\line(0,1){5}}
\put(100,40){\vector(1,1){20}}
\put(120,60){\line(1,1){20}}
\put(140,80){\vector(-1,0){40}}
\put(100,80){\line(-1,0){40}}
\put(60,80){\vector(1,-1){20}}
\put(80,60){\line(1,-1){20}}
\multiput(60,80)(-15,15){2}{\line(-1,1){10}}
\multiput(140,80)(15,15){2}{\line(1,1){10}}
\put(85,10){$Q$}
\put(110,35){$\gamma_\alpha\gamma^5$}
\put(150,80){$\gamma_\nu\gamma^5$}
\put(20,80){$\gamma_\mu\gamma^5$}
\put(50,95){$p$}
\put(145,95){$p'$}
\put(60,57){$q$}
\put(125,55){$q'$}
\end{picture}}
\ec                         %instead of \end{center}
\vspace{-2mm}
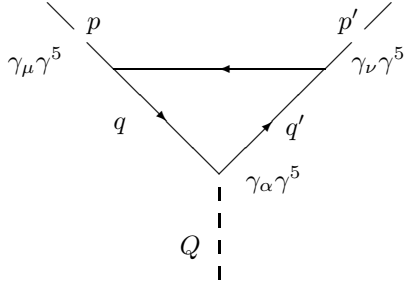
\caption{Triangle graph occurring in the axial-vector $\mathcal{PT}$-symmetric
quantum electrodynamics.}
\label{fig1}
\efg                        %instead of \end{figure}
This would seem to completely change the weak coupling expansion from that
in normal QED.  In particular, $F_1$ is not simply obtained from that in 
ordinary QED!

It is well-known that the AAA triangle graph possesses 
an axial-vector anomaly.
This is usually seen as a consequence of enforcing Bose symmetry,
which thereby resolves the ambiguity associated with a superficially
linearly divergent loop integration \cite{gj72}.
This is in contrast with the more
familiar AVV graph, where the axial anomaly arises from enforcement of
vector current conservation \cite{triangle}.

\subsection{Triangle Anomaly}
It should be instructive to see how this comes about in a method in which
all quantities are explicitly finite.  This is the 
``causal'' or ``dispersive'' approach,
advocated by Schwinger's source-theory group in 1970s.  As it happens,
I have an unpublished manuscript \cite{ucla}
 which presents the calculation of precisely the above graph in spectral
form, for the general situation in which all particles have masses:
\begin{eqnarray}
I_{\mu\nu\alpha}&=&Q_\alpha\epsilon_{\mu\nu\lambda\sigma}p^\lambda p^{\prime
\sigma}\int_{4m^2}^\infty \frac{dM^2}{2\pi {\rm i}}\frac{A_2(M^2)}
{M^2+Q^2-i\epsilon}
\nonumber\\
&&\mbox{}+\epsilon_{\mu\nu\lambda\alpha}(p-p')^\lambda
\int_{4m^2}^\infty\frac{dM^2}{2\pi {\rm i}}\frac{A_1(M^2)}{M^2+Q^2-i\epsilon}.
\end{eqnarray}
Here $m$ is the mass of the fermion, 
the outgoing ``photon'' momenta are on the mass shell,
$p^2=p^{\prime2}=-\mu^2$, and we
have set $p^\mu\to0$ and $p^{\prime\nu}\to 0$ appropriate for real, outgoing
vector particles. 
This is a consequence of the property of the three polarization vectors for a
massive, spin-1 particle:
\begin{equation}
e^\mu_{p\lambda}:\quad p_\mu e^\mu_{p\lambda}=0,\quad\sum_{\lambda=1}^3
e^\mu_{p\lambda}e^\nu_{p\lambda}{}^*=g^{\mu\nu}+\frac{p^\mu p^\nu}{\mu^2}.
\end{equation}

 The spectral functions are determined by taking a cut across
the $q$-$q'$ lines in the graph: ($M^2=-Q^2$)
\begin{eqnarray}
\tilde I_{\mu\nu\alpha}&=&\int d\omega_q\,d\omega_{q'}(2\pi)^4\delta(Q-q-q')
\frac1{m^2+(p-q)^2}\nonumber\\
&&\times\mbox{Tr}\,\bigg[\gamma_5\gamma_\mu[m+\gamma(p-q)]\gamma_5\gamma_\nu
(m-\gamma q')\gamma_5\gamma_\alpha(m+\gamma q)\bigg]\nonumber\\
&=&Q_\alpha\epsilon_{\mu\nu\lambda\sigma}p^\lambda p^{\prime\sigma} A_2(M^2)
+\epsilon_{\mu\nu\lambda\alpha}(p-p')^\lambda A_1(M^2).
\label{cut}
\end{eqnarray}
Here the invariant phase-space measure is
\begin{equation}
d\omega_p=\frac{(d{\bf p})}{(2\pi)^3}\frac1{2p^0}.
\end{equation}
Explicit formula may be straightforwardly worked out for the spectral functions
for this general mass case:
\begin{eqnarray}
A_1(M^2)&=&-\frac1{16\pi}\frac{v}{\zeta^4}\bigg\{3-4\zeta^2+\zeta^4\nonumber\\
&&\quad\mbox{}-\bigg[3-\zeta^2-3\zeta^4+\zeta^6
+4\zeta^2v^2(1-\zeta^2)
\bigg]\frac1{4v\zeta}\ln\phi\bigg\},
\end{eqnarray}
\begin{eqnarray}
A_2(M^2)&=&-\frac1{8\pi M^2}\frac{v}{\zeta^4}\bigg\{3-4\zeta^2+\zeta^4\nonumber
\\
&&\quad\mbox{}-\bigg[3-\zeta^2+5\zeta^4+\zeta^6
-4\zeta^2v^2(1+\zeta^2)
\bigg]\frac1{4v\zeta}\ln\phi,\bigg\}.
\end{eqnarray}
Here
\begin{equation}
v=\left(1-\frac{4m^2}{M^2}\right)^{1/2},\quad
\zeta=\left(1-\frac{4\mu^2}{M^2}\right)^{1/2},
\end{equation}
and
\begin{equation}
\phi=\frac{1+\zeta^2+2v\zeta}{1+\zeta^2-2v\zeta}.
\end{equation}

The anomaly is obtained by taking the divergence with respect to
the unrestricted vertex $\alpha$, that is, by multiplying by $Q^\alpha$.
\begin{eqnarray}
{\rm i}Q^\alpha I_{\mu\nu\alpha}=\epsilon_{\mu\nu\lambda\sigma}p^\alpha p^{\prime
\sigma}\left\{\int_{4 m^2}^\infty \frac{dM^2}{2\pi}\frac{2 A_1(M^2)-
M^2 A_2(M^2)}{M^2+Q^2-i\epsilon}+a\right\},\end{eqnarray}
where the spectral function in the integral is just the $Q^\alpha$ contraction
with the cut amplitude (\ref{cut}). It has a rather simple form:
\begin{equation}
2A_1-M^2A_2=-\frac1{4\pi\zeta^3}(\zeta^2-v^2)\ln\phi.
\end{equation}
The integral vanishes as $m\to0$.
The anomaly arises because now $Q^2\ne -M^2$:
\begin{equation}
a=\int_{4m^2}^\infty \frac{dM^2}{2\pi}A_2(M^2)=\frac1{(2\pi)^2}.
\end{equation}
The evaluation, independent of $\mu^2/m^2$, may be straightforwardly
verified. 
 Of course, the case of direct interest is much simpler, because
$\mu=0$ for a photon:
\begin{equation}
a=\frac1{4\pi^2}\int_0^1 v\,dv\,\ln\left(\frac{1+v}{1-v}\right)
=\frac1{4\pi^2}.
\end{equation}
It might appear that the correct limit here is to first set $m=0$, then
take $\mu\to0$.  However this is an  anomalous threshold
 situation, best handled by analytic continuation from the $m>\mu$ case.

So the $\mathcal{PT}$-symmetric electrodynamics proposed by us in 
Ref.~\cite{bm99} seems to possess a serious difficulty: 
Because $\mathcal{P}$-violating
 Green's functions occur, containing an odd number
of $\gamma^5\gamma^\mu$ vertices---that is, there is no Furry's theorem---an
axial vector anomaly can occur in the theory.  We have explicitly calculated
such an anomaly.  Therefore, it appears that the theory is rendered 
nonrenormalizable.
The nonrenormalizable divergent graphs appear when, for example, the
two photons of the triangle graph are attached to a fermion line, as shown
in Fig.~\ref{fig2}.

\bfg[t]                     %instead of \begin{figure}[t]
\bc                         %instead of \begin{center}
\mbox{\begin{picture}(200,120)
\multiput(100,0)(0,10){4}{\line(0,1){5}}
\put(100,40){\vector(1,1){20}}
\put(120,60){\line(1,1){20}}
\put(140,80){\vector(-1,0){40}}
\put(100,80){\line(-1,0){40}}
\put(60,80){\vector(1,-1){20}}
\put(80,60){\line(1,-1){20}}
\multiput(60,80)(-15,15){2}{\line(-1,1){10}}
\multiput(140,80)(15,15){2}{\line(1,1){10}}
\put(-20,110){\vector(1,0){120}}
\put(100,110){\line(1,0){120}}
\end{picture}}
\ec                         %instead of \end{center}
\vspace{-2mm}
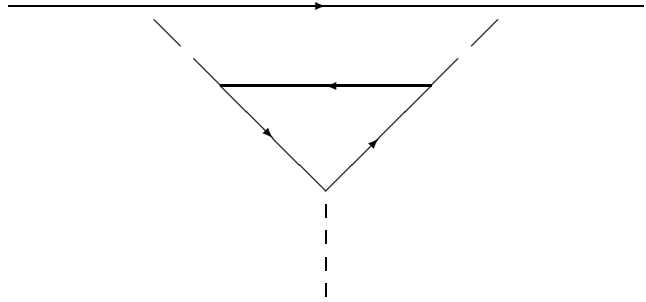
\caption{Divergent graph which cannot be renormalized in axial-vector
electrodynamics.}
\label{fig2}
\efg                        %instead of \end{figure}

There are a number of caveats that must be noted concerning this conclusion:
\begin{itemize}
\item The Feynman rules, and the unitarity arguments that lead to the
dispersion relations, may have to be modified in the $\mathcal{PT}$ theory
because the contours that define the theory do not lie along the real axis
in general.
\item There may be a subtlety associated with massless fermions.
\item Although this particular version of $\mathcal{PT}$QED may be flawed,
it may be possible to find other variants.  (We shall suggest such
a possibility in the next section.)
\item Of course, it must always be acknowledged that the criterion of
renormalizability is not the final arbiter; nonrenormalizable theories
can still be useful, effective ones.

\end{itemize}
\section{Alternative $\mathcal{PT}$-symmetric electrodynamics}

The axial-vector current theory seems to be fatally flawed.  Fortunately,
there is an alternative $\mathcal{PT}$-symmetric theory which seems to
be, in fact, closer to the spirit of the general development.  Instead,
I propose using the usual current,
\be
j^\mu=\frac12\psi\gamma^0\gamma^\mu eq\psi,
\ee
but couple it to an {\em axial-vector\/} photon field $A_\mu$:
\be
\mathcal{L}_{\rm int}={\rm i}j^\mu A_\mu.
\ee
Note that the factor of $\rm i$ is inserted to ensure $\mathcal{PT}$ invariance.
Under either $\mathcal{P}$ or $\mathcal{T}$ separately, the time component
of $j^\mu$ does not change, while the space component reverses sign, hence
\begin{equation}
\mathcal{PT}:\quad j^\mu\to j^\mu.
\end{equation}
This is consistent with the modified Maxwell equations,
\begin{equation}
\mbox{\boldmath{$\nabla$}}\cdot {\bf E}={\rm i}\rho,\quad
 \mbox{\boldmath{$\nabla$}}\times{\bf B}=\frac\partial{\partial t}{\bf E}+
 {\rm i}{\bf j},
 \end{equation}
 provided under $\mathcal{PT}$:\footnote{There actually seem to be
 two possible theories: Either $\bf E$ is an axial vector and $\bf B$ is
 a polar vector, if the scalar potential but not the vector potential,
 changes sign under parity (this smells a bit like magnetic charge),
 or $\bf B$ is an axial vector and $\bf E$ is a polar vector if the
 vector potential not the scalar potential changes sign under parity.
 The latter is more a theory of electric charge.}
 \be
 \mathcal{PT}:\quad {\bf E}\to{\bf E},\quad {\bf B}\to{\bf B},\quad
 A^{\mu}\to- A^{\mu}.  
\ee
 $\mathcal{PT}$ invariance of the theory is thus assured:
 \be
 \mathcal{L}_{\rm int}\to \mathcal{L}_{\rm int}.
 \ee
This theory seems to be a $\mathcal{PT}$QED with
\begin{itemize}
\item Furry's theorem holding (no odd Green's functions)
\item No axial-vector anomaly
\item Usual perturbation theory with $\alpha\to-\alpha$.
\end{itemize}
Everything we erroneously had said about the ${\rm i}j_5^\mu A_\mu$ theory
does seems to hold for the ${\rm i}j^\mu A_\mu$ theory.

This new theory would seem to be asymptotically free, since the sign
of the beta function reverses from that in ordinary QED.  
An antiscreening effect may occur here because of the attraction
of like charges.
Quantum-mechanically, the sign of the vacuum polarization reverses.
Whether this implies confinement is under investigation.

We are also examining questions of the stability of the vacuum, in the
presence of a strong electric field $E$.  The Schwinger mechanism says that
the probability of pair creation in ordinary QED is proportional to
\be
P(0\to e^+e^-)\sim {\rm e}^{-\pi m^2/eE};
\ee
what happens in our case?  

We will be examining this new ``conventional $\mathcal{PT}$-symmetric''
QED for consistency, and we ask whether somewhere  might it
be realized in nature?

\bigskip
{\small I thank Qinghai Wang for extensive discussions, and the US
Department of Energy for financial support.  I am very grateful for Miloslav
Znojil for inviting me to participate in this most exciting workshop.
Carl Bender and I thank Howard E. Brandt for detecting the error
in the conventional eigenvalues of $F_1$ in Ref.~\cite{bm99}.}
                 %for \end{thebibliography}

\end{document}